\documentclass[a4paper,11pt]{article}
\usepackage{jheppub} 
\usepackage{lineno}
\usepackage{xcolor}
\usepackage[T1]{fontenc} 
\usepackage{amsmath, amssymb, amsfonts}
\usepackage{textcomp,amsmath,amssymb} 
\usepackage{scalerel}
\usepackage{mathtools}
\usepackage{dsfont}
\usepackage{float}
\usepackage{subcaption}
\usepackage{graphicx}   
\usepackage{xcolor}

\title{\boldmath  Quantum Estimation Theory Limits in Neutrino Oscillation Experiments}
\author{Claudia Frugiuele,  Marco G. Genoni, Michela Ignoti,  Matteo G. A. Paris}
\affiliation{INFN Sezione di Milano, I-20133 Milano, Italy}
\affiliation{Dipartimento di Fisica {\em Aldo Pontremoli}, Universit\`a degli Studi di Milano, I-20133 Milano, Italy}
\emailAdd{claudia.frugiuele@mi.infn.it}
\abstract{ 

Measurements of the Pontecorvo–Maki–Nakagawa–Sakata (PMNS) neutrino mixing parameters have entered a precision era, enabling increasingly stringent tests of neutrino oscillations. Within the framework of quantum estimation theory, we investigate whether flavor measurements, 
the only observables currently accessible experimentally, are optimal for extracting the oscillation parameters.
We compute the Quantum Fisher Information (QFI) and the classical Fisher Information (FI) associated with ideal flavor projections for all oscillation parameters, considering accelerator muon (anti)neutrino and reactor electron antineutrino beams propagating in vacuum.
Two main results emerge. First, flavor measurements saturate the QFI at the first oscillation maximum for $\theta_{13}$, $\theta_{23}$, and $\theta_{12}$, demonstrating their information-theoretic optimality for these parameters. In contrast, they are far from optimal for $\delta_{CP}$. In particular, only a small fraction of the available information on $\delta_{CP}$ is extracted at the first maximum; the sensitivity improves at the second maximum, in line with the strategy of ESS$\nu$SB, a planned facility.
Second, the QFI associated with $\delta_{CP}$ is approximately one order of magnitude smaller than that of the mixing angles, indicating that the neutrino state intrinsically encodes less information about CP violation. Nevertheless, this quantum bound lies well below current experimental uncertainties, implying that the present precision on $\delta_{CP}$ is not fundamentally limited.
Our results provide a quantitative framework to disentangle fundamental from practical limitations and establish a benchmark for optimizing future neutrino facilities.
}

\begin{document}
\maketitle
\section{Introduction}
\label{sec:intro}

Neutrino oscillations, described by the Pontecorvo–Maki–Nakagawa–Sakata (PMNS) matrix, provide a powerful framework for probing fundamental neutrino properties, including leptonic CP violation encoded in the phase $\delta_{CP}$. 
Despite these achievements, $\delta_{CP}$ remains poorly constrained compared to the other oscillation parameters, with current global fits providing only weak indications of CP violation \cite{Esteban_2024}. The NuFIT~v6.0 global analysis \cite{Esteban_2024} reports, for the normal ordering (NO) scenario, $\delta_{CP}^{(\mathrm{NO})} = 177^{\circ}\,^{+19^{\circ}}_{-20^{\circ}}$, based primarily on data from the long-baseline accelerator experiments T2K~\cite{Abe2023} and NO$\nu$A~\cite{PhysRevD.106.032004} while, for example, for the mixing angle $\theta_{13} = 8.52^{\circ}\,^{+0.11^{\circ}}_{-0.11^{\circ}}$, highlighting the current experimental challenges in probing CP violation in the neutrino sector compared to the others oscillation parameters. Analogous results are obtained under the inverse ordering (IO) hypothesis. Next-generation facilities such as DUNE~\cite{dune,universe10050221}, T2HK~\cite{yokoyama2017hyperkamiokandeexperiment} and ESS$\nu$SB ~\cite{Alekou_2023, Alekou2021}  aim to significantly improve this precision and may provide the first definitive evidence of leptonic CP violation.
The estimation of the PMNS parameters can be naturally formulated within the framework of quantum estimation theory (QET)~\cite{helstrom1976quantum,Holevo2011b,paris2009,Albarelli2019c} by treating the neutrino flavor states $|\nu_\alpha(t)\rangle$  as parameter-dependent quantum states whose evolution, governed by the PMNS matrix, encodes information about the oscillation parameters. The sensitivity to these parameters can then be quantified in terms of the classical Fisher Information (FI) and the Quantum Fisher Information (QFI), which set the ultimate bounds on their attainable precision via the Cramér–Rao bound.
\par
In conventional QET, the main goal is to identify or engineer optimal quantum states and measurement schemes that saturate the Cramér–Rao bound, thereby achieving the ultimate precision allowed by quantum mechanics. In neutrino physics, however, the situation is fundamentally different: the quantum state and the measurement basis are fixed by nature. The neutrino flavor states are uniquely determined by the PMNS matrix, and the detection process effectively performs a flavor measurement. Consequently, the central question is not only what is the ultimate precision limit imposed by the neutrino  flavor state itself, but also whether the unavoidable restriction to flavor measurements constitutes an additional intrinsic limitation on the achievable precision for specific oscillation parameters. This is precisely the question we address in this work by comparing the QFI and the FI associated with flavor measurements across all neutrino oscillation parameters.
\par
We emphasize that the QFI is not merely a theoretical abstraction. It provides the ultimate precision bound set by the neutrino quantum state itself, independent of any specific detection scheme. Comparing the QFI with the classical FI of flavor measurements reveals whether a given experimental configuration (baseline, energy, channel) is information-theoretically optimal or whether room for improvement remains. This distinction, that between fundamental quantum limits and practical measurement choices, is inaccessible to standard $\chi^2$ or likelihood analyses, which always assume a fixed measurement strategy. Thus, QET offers a unique and rigorous tool for optimizing future neutrino facilities, as demonstrated here for the second oscillation maximum and for the intrinsic hierarchy between $\delta_{CP}$ and the mixing angles.

\par
This study extends the analysis of Ref.~\cite{prl}, which focused exclusively on $\delta_{CP}$, to all neutrino oscillation parameters, enabling a systematic comparison of their achievable estimation precision.
This work is part of a broader program aimed at applying quantum information tools~\cite{Fedida2023,Fedida2024,Afik:2025ejh}, and in particular of QET~\cite{bak2016information,lashkari2016canonical,trivella2017holographic,QETcosmology2018,QETcosmology2020,erdmenger2020information,QETcosmology2021,feng2022quantum,candeUGC, seveWEP, cepoTD,AlessioQED,AlessioCosmological,collider,nogueira2016quantumestimationneutrinooscillations}, to high-energy physics.\\

\par
The present paper is structured as follows. We first give a brief overview on QET and  describe how this formalism can be adapted to neutrino oscillations and specify the assumptions underlying our analysis in Section \ref{Problem Modelization}. We then compute the QFI in Section \ref{Quantum Fisher Information} and the FI in Section \ref{Fisher Information} for all oscillation parameters, providing a systematic comparison of their sensitivities. Section \ref{conclusions} presents our conclusions and outlook.

\section{Problem Modelization} \label{Problem Modelization}
In this section, we present the basic framework of QET~\cite{helstrom1976quantum,Holevo2011b,paris2009,Albarelli2019c} and describe how we can describe the estimation of the parameters involved in a neutrino oscillations' experiment as a quantum metrology protocol,  outlining all the assumptions made to simplify our study.
\subsection{Quantum estimation theory in a nutshell}\label{s:QET}
Let us consider a physical parameter $\lambda$ we want to estimate and a (classical) measurement that is mathematically described by the conditional probability $p(x|\lambda)$ of observing the classical outcome $x$ given the value of the parameter $\lambda$. By repeating the experiment $M$ times one accumulates a set of measurement outcomes ${\bf x}=\{x_1,\dots,x_M\}$. An estimator $\tilde{\lambda}$ is defined by a map from the measurement outcomes ${\bf x}$ to the possible values of the parameter $\lambda$. 
The achievable precision of any unbiased estimator $\tilde{\lambda}$ is limited by the Cramér–Rao bound,
\begin{equation}
\mathrm{Var}_{\tilde{\lambda}}(\lambda) \ge \frac{1}{M F(\lambda)},
\label{cramerrao}
\end{equation}
where $F(\lambda)$ is the FI, which is a function of the conditional probability $p(x|\lambda)$, 
\begin{align}
    F(\lambda) = \sum_x p(x|\lambda) (\partial_\lambda \log p(x|\lambda))^2 \,.
    \label{eq:FI}
\end{align}
Let us now move to the quantum scenario. As in the following we will assume that our experiments can be efficiently described by pure states, that is normalized vectors in a Hilbert space, but a more general framework in terms of density operators can be equivalently provided~\cite{Helstrom:1969fri,paris2009}. We start from an initial quantum state $|\psi_0\rangle$, which does not depend on $\lambda$, and apply a unitary transformation $U_\lambda$ that encodes the parameter $\lambda$ on the output state $|\psi_\lambda \rangle= U_\lambda |\psi_0\rangle$. We can now assume that a quantum measurement, described by a projector-valued-measure (PVM) and thus to a complete set of projectors on orthonormal quantum states $
\{\Pi_x = |x\rangle\langle x|\}$ is performed (also in this case, one can extend the framework including the more general positive-operator-valued-measures, POVMs). The aforementioned conditional probability, describing the quantum experiment, can be directly obtained via the Born rule, as $p(x|\lambda) = |\langle x | \psi_\lambda\rangle|^2$, allowing to evaluate the corresponding FI via the formula provided above in Eq.~\eqref{eq:FI}. Remarkably, in the quantum case one can obtain a more fundamental bound, dubbed quantum Cram\'er-Rao bound, such that
\begin{equation}
\mathrm{Var}_{\tilde{\lambda}}(\lambda) \ge \frac{1}{M F(\lambda)}
\geq \frac{1}{M H(\lambda)}  \,,    \label{eq:qcrb}
\end{equation}
where $H(\lambda)$ denotes the QFI and corresponds to the maximum of the FIs over all the possible measurements allowed by quantum mechanics. For experiments described by pure states $|\psi_\lambda\rangle$ the QFI can be evaluated via the formula~\cite{Helstrom:1969fri,paris2009}
\begin{equation}
\begin{split}
  H(\lambda) 
  &= \max_{\Pi_x} F(\lambda) \nonumber \\
  &= 4\left(
     \langle\partial_\lambda \psi_\lambda|\partial_\lambda \psi_\lambda\rangle
     - \bigl|\langle\psi_\lambda|\partial_\lambda \psi_\lambda\rangle\bigr|^2
     \right).
\end{split}
\label{eq:qfi_stati puri}
\end{equation}
Hence, while the FI $F(\lambda)$ quantifies the amount of information on the parameter $\lambda$ given the quantum state $|\psi_\lambda\rangle$ and a specific measurement $\{\Pi_x\}$, the QFI $H(\lambda)$ represents the maximum amount of information that can, in principle, be extracted from the quantum state regardless of the specific measurement strategy.\par

\subsection{Quantum estimation theory applied to Neutrino Physics}
We now focus on neutrinos and how to apply the formalism just outlined to the study of the oscillation parameters. Neutrinos are among the most elusive particles in nature, as they interact exclusively through the weak interaction. Although the Standard Model predicts massless neutrinos, the experimental observation of neutrino oscillations demonstrates that they possess nonzero masses. Neutrino oscillations in vacuum are a quantum mechanical phenomenon made possible by the existence of non-degenerate neutrino masses ($m_1,m_2,m_3$) and lepton flavor mixing.
Flavor mixing arises in the lepton sector because the orthonormal basis of interacting flavor eigenstates  ($|\nu_e\rangle,|\nu_\mu\rangle,|\nu_\tau\rangle$) does not coincide with the orthonormal basis of mass eigenstates ($|\nu_1\rangle,|\nu_2\rangle,|\nu_3\rangle$). The left-handed neutrino fields in the flavor basis and those in the mass basis are related by a unitary transformation described by the Pontecorvo-Maki-Nakagawa-Sakata (PMNS) matrix. Considering neutrinos as Dirac fermions, the PMNS matrix can be parametrized by three mixing angles ($\theta_{12},\theta_{13},\theta_{23}$) and one Dirac CP-violating phase $\delta_{CP}$:

\begin{equation}
        \resizebox{\textwidth}{!}{$
U=\begin{pmatrix}
 \cos{\theta_{12}} \cos{\theta_{13}} & \sin{\theta_{12}} \cos{\theta_{13}} & e^{-i \delta_{CP} } \sin{\theta_{13}} \\
 -\sin{\theta_{12}}\cos{\theta_{23}}-e^{i \delta_{CP}} \cos{\theta_{12}} \sin{\theta_{13}} \sin{\theta_{23}} & \cos{\theta_{12}} \cos{\theta_{23}}-e^{i \delta_{CP}} \sin{\theta_{12}} \sin{\theta _{13}} \sin{\theta_{23}} & \cos{\theta_{13}} \sin{\theta_{23}} \\
 \sin{\theta_{12}} \sin{\theta_{23}}-e^{i \delta_{CP}} \cos{\theta_{12}} \sin {\theta_{13}} \cos{\theta_{23}} & -\cos{\theta_{12}} \sin{\theta_{23}}-e^{i \delta_{CP}} \sin{\theta_{12}} \sin{\theta_{13}} \cos{\theta_{23}} & \cos{\theta_{13}} \cos{\theta_{23}} \\
\end{pmatrix}
$}.
    \end{equation}
    
Let us consider an initial beam of definite flavor $\nu_\alpha$. The quantum state $|\nu_\alpha\rangle$ can then be  expressed in terms of mass eigenstates via the PMNS matrix:
\begin{equation}
    |\nu_{\alpha}\rangle=\sum_i U_{\alpha i}^* |\nu_i\rangle  \quad \text{with} \quad \alpha=e,\mu,\tau,
    \label{eq: stato 0}
\end{equation}
where $|\nu_i\rangle$ denotes the mass eigenstate with eigenvalue $m_i$. Therefore a neutrino produced via weak interactions is a coherent superposition of mass eigenstates. In vacuum each mass eigenstate $|\nu_i\rangle$ is an eigenstate of the free Hamiltonian with eigenvalue $E_i=\sqrt{p^2+m_i^2}$, where p is the momentum of the neutrino beam and  we assume the same momentum for all mass eigenstates:
\begin{equation}
    |\nu_i(t)\rangle=e^{-i E_it}|\nu_i\rangle.
\end{equation}
Thus propagation modifies the original coherent superposition in Eq. \eqref{eq: stato 0}, which is no longer a pure flavor eigenstate. If the neutrino is generated at time $t_0=0$ in a flavor eigenstate $|\nu_{\alpha}\rangle$, its state at a generic time $t$ can be written as:
\begin{equation}
    |\nu_\alpha(t)\rangle=\sum_i U_{\alpha i}^* e^{-i E_it} |\nu_i\rangle=\sum_i U_{\alpha i}^* e^{-i E_it} \sum_\beta U_{\beta i} |\nu_\beta\rangle,
    \label{eq:change_basis}
\end{equation}
where in the second equality the state has been rewritten in the flavor basis.
From this expression we observe that the neutrino state at time $t$ depends on the parameters of the PMNS matrix, as well as on $E_i$ and $t$.\par
The probability that a neutrino produced with flavor $\alpha$ is detected at time $t$ with flavor $\beta$ is therefore given by:
\begin{equation}
\begin{split}
    P(|\nu_\alpha\rangle \to |\nu_\beta\rangle)&= |\langle\nu_\beta|\nu_\alpha(t)\rangle|^2=\left|\sum_i U_{\beta i}U_{\alpha i}^* e^{-iE_it}\right|^2.
    \label{eq : prob oscil}
    \end{split}
\end{equation}
Since neutrinos are ultra relativistic, the approximation  $E_i\approx p+m_i^2/2E$ with $E\approx p$ is valid and the probability can be rewritten as:
\begin{equation}
\begin{split}
    P(|\nu_\alpha\rangle \to |\nu_\beta\rangle)&=\sum_{i,j} U_{\beta i}U_{\alpha i}^* U_{\beta j}^*U_{\alpha j} e^{i\frac{\Delta m_{ji}^2 L}{2E}},
    \end{split}\label{eq : prob}
\end{equation}
where $\Delta m_{ij}^2 \equiv m_j^2-m_i^2$ and $L\approx ct$. Thus the oscillation probability depends on the PMNS parameters, on the squared mass differences $\Delta m_{ij}^2$ and on the ratio $L/E$ between the distance traveled by the neutrino and its energy. In experiments $L$ is referred to as the baseline, i.e the distance between the production and detection points of the neutrino beam.\par
Expanding Eq. \eqref{eq : prob} we obtain:
\begin{align}
\begin{split}
    P(|\nu_\alpha\rangle \to |\nu_\beta\rangle)&=\delta_{\alpha\beta} \\&-4 \sum_{i<j} \text{Re}[U_{\alpha i}U_{\beta i}^*U_{\alpha j}^*U_{\beta j}]\sin^2\biggl(\frac{\Delta m_{ji}^2 L}{4E}\biggl)\\&+ 2 \sum_{i<j} \text{Im}[U_{\alpha i}U_{\beta i}^*U_{\alpha j}^*U_{\beta j}]\sin\biggl(\frac{\Delta m_{ji}^2 L}{2E}\biggl).
    \end{split}\label{eq : prob explicit}
\end{align}
The probability of flavor change is a sum of sinusoidal and sine squared functions in $L/E$, so it necessarily oscillates with $L/E$. For antineutrino oscillations $U$ must be replaced by $U^*$ so that the imaginary part in Eq. (\ref{eq : prob explicit}) changes sign.

\par
\begin{figure}[h]
    \centering
    \includegraphics[width=0.9\textwidth]{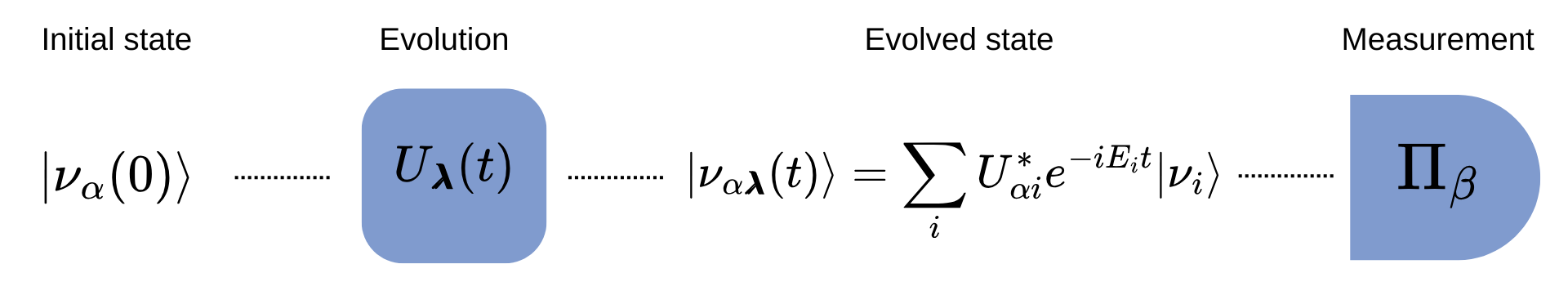}
    \caption{Neutrino oscillations as a quantum metrology protocol: a neutrino with a specific flavor $\alpha$ is prepared at time $t=0$. After evolving for a time $t$ the neutrino quantum state can be written as a superposition of flavor eigenstates, with coefficients that depends on the different parameters ${\bf \lambda}$ that characterize the neutrino oscillations. A final measurement (typically corresponding to a flavor detection) is performed on the output state.}
    \label{fig:schema_misura_2}
\end{figure}
We now recast the neutrino oscillation experiments described above within the quantum metrology framework introduced in Sec.~\ref{s:QET}. The scenario is schematically illustrated in Fig.~\ref{fig:schema_misura_2}. The initial (anti)neutrino quantum state is a flavor eigenstate \( |\psi_0\rangle = |\nu_\alpha\rangle \); after an evolution time \( t \), the neutrino is then described by the quantum state \( |\psi_\lambda\rangle = |\nu_\alpha(t)\rangle \), given in Eq.~\eqref{eq:change_basis}, which explicitly depends on all the parameters characterizing the experiment, and a final measurement is performed.

In the following, we focus on the estimation of the PMNS matrix parameters, i.e.  we consider the parameters set 
$
\lambda = \{ \delta_{CP}, \theta_{12}, \theta_{13}, \theta_{23} \},
$ and analyze their achievable estimation precision.
Although a full multiparameter analysis could in principle be performed~\cite{Albarelli2019c}, we 
focus to the single-parameter scenario, given  the fundamental nature of the questions motivating our study. Accordingly, we consider the estimation of a single parameter at a time, assuming all remaining parameters to be known and fixed to their NuFit~6.0 best-fit values~\cite{Esteban_2024}.
In particular, we assume the Normal Ordering (NO) for the neutrino mass hierarchy. The results remain analogous under the Inverted Ordering (IO) hypothesis, since for vacuum oscillations the information encoded in the quantum state and the corresponding observables are only weakly dependent on the mass ordering.\par
By computing and comparing the QFI and the FI we assess whether current experimental uncertainties are dominated by statistical fluctuations or are already approaching the intrinsic quantum bounds set by the structure of neutrino oscillations.
 
\subsection{Experimental framework and assumptions}
In our analysis, we aim to determine whether the necessity of performing flavor measurements and the impossibility of engineering new neutrino quantum states introduce an additional intrinsic quantum limit on measurement precision, beyond that set by the QFI.
To this end, we make a set of assumptions regarding both the neutrino quantum state and its detection. Regarding the state, as we have described just above, the mass eigenstates $|\nu_i\rangle$ are modeled as plane waves propagating in vacuum ~\cite{Akhmedov:2009rb, Eliezer:1975ja, Bilenky:1976yj, Fritzsch:1975rz}. This idealization allows for a clean evaluation of the maximal information encoded in the oscillation dynamics. A realistic treatment including matter effects, wave-packet descriptions of mass eigenstates (and the associated decoherence), would generally reduce the achievable precision. However, such refinements are beyond the scope of the present analysis.
As regards the detection, we  adopt the concept of an \emph{ideal flavor measurement}, the direct detection of neutrino flavor without experimental refinements such as energy resolution, baseline smearing, detector inefficiencies, or backgrounds.
Accordingly, we model each experiment’s neutrino beam as monochromatic and characterized by its peak energy, allowing us to isolate the intrinsic information content arising solely from oscillation dynamics. We focus on accelerator and reactor experiments and assume a detector capable of perfectly identifying neutrino flavor with exact knowledge of the energy $E$ and baseline $L$. As observed in Eq. \eqref{eq : prob}, the oscillation probability evolves with the ratio $L/E$. Since we are interested in quantifying the amount of information on the PMNS parameters at the different oscillation peaks, our results will be shown as a function of the ratio $L/E$. In the following we will provide more details characterizing accelerator and reactor experiments.

\begin{itemize}
     \item \textbf{Accelerator Experiments}: In accelerator neutrino experiments, the initial beam consists predominantly of $\nu_\mu$ or $\overline{\nu}_\mu$, depending on whether the experiment operates in neutrino or antineutrino mode. These experiments are characterized by a near-detector (ND), where the initial interaction rate of neutrinos is measured and a far-detector (FD) where the oscillations $\nu_\mu(\overline{\nu}_\mu) \to \nu_e(\overline{\nu}_e)$ and disappearance $\nu_\mu(\overline{\nu}_\mu) \to \nu_\mu(\overline{\nu}_\mu)$ are measured.

We evaluate the QFI and classical FI for experiments that can be effectively treated as operating in vacuum, namely T2K/T2HK~\cite{T2K:2011qtm}\cite{PhysRevD.106.073006} and ESS$\nu$SB~\cite{Alekou_2023, Alekou2021}. The parameters of interest are $\delta_{CP}$, $\theta_{23}$, and  $\theta_{13}$.
 We focus on the peak neutrino energies, $E_\nu = 0.6$~GeV for T2K/T2HK and $E_\nu \approx 0.25$~GeV for ESS$\nu$SB, with corresponding baselines of 295~km and 360~km. Matter effects are neglected in this analysis; consequently, experiments where these are significant, such as NO$\nu$A and DUNE, are excluded. 
Furthermore, accelerator experiments typically analyze neutrino and antineutrino oscillations separately. As a result, the QFI and FI contributions from neutrinos and antineutrinos can be treated as originating from two independent datasets, which allows for a straightforward combination of their sensitivities in the estimation of the oscillation parameters, as we will describe in the next subsection. 
   \item \textbf{Reactor Experiments}:  For reactor experiments, the neutrino beam consists of $\overline{\nu}_e$ produced by fission reactions in nuclear reactors. The analysis focuses on $\theta_{13}$ for Daya Bay \cite{An_2014} and on $\theta_{12}$ for KamLAND \cite{PhysRevLett.90.021802}. Both experiments study the oscillation channel $\overline{\nu}_e \to \overline{\nu}_e$ and Daya bay also presents near detectors. The peak neutrino energies used to evaluate the FI and QFI are approximately $E_\nu \simeq 3.6$~MeV for KamLAND and $E_\nu \simeq 3$~MeV for Daya Bay, with corresponding baselines of 180~km and 1.64~km, respectively.
\end{itemize}

\subsection{Estimation of total number of events}
\label{s:numberevents}
Particular care must be devoted to the number of independent experimental runs ($M$) entering the Cramér--Rao bound~\eqref{eq:qcrb}. In our scenario, ($M$) corresponds to the so-called {\it non-oscillated} (anti)neutrinos, that is, the total number of detected neutrinos in the absence of oscillation phenomena, formally obtained by assuming that the PMNS matrix is the identity. By an abuse of terminology, we will refer to (M) as the number of events; however, as mentioned above and as will be explained in more detail below, this number should not be confused with the number of neutrinos detected at the far detector.
In the results discussed in the following sections, we will consider approximate values of $M$, corresponding to the different experiments considered. Here we will describe how this information can be obtained from the physical parameters describing such experiments.\\

We start by considering accelerator experiments where a muon neutrino beam is generated (the arguments can be readily extended to the case of antineutrinos by properly changing the different quantities entering in the formulas). In these experiments, one typically employs two detectors: a Near Detector (ND) and a Far Detector (FD). The number of events at the ND for a beam of muon neutrinos $\nu_\mu$ is given by:
\begin{equation}
    M_{\nu_\mu}^{\small ND}=\phi^{\small ND}_{\nu_\mu} \ \sigma_{CC}^{ND} \ n_{targ}^{\small ND} \ \eta_\mu
    \label{eq : eventi ND}
\end{equation}
where $\phi_{\nu_\mu}^{\small ND}$ is the flux of $\nu_\mu$ at the ND, $\sigma_{CC}^{ND}$ is the cross section for charged current interaction at the ND, $n_{targ}^{\small ND}$ is the number of targets in the detector and $\eta_\mu$ is the detector efficiency for the detection of the charged lepton associated with $\nu_\mu$. 

As mentioned above the FD probes either the appearance or the disappearance channel. As a consequence the number of events at the FD for neutrinos of  flavors $\beta$ ($\beta=e$ for the appearance channel, and $\beta=\mu$ for the disappearance channel) is given by:
\begin{equation}
    M_{\nu_\beta}^{FD}=\phi_{\nu_\mu}^{FD} \ \sigma_{CC}^{FD} \ n_{targ}^{FD} \ \eta_\beta \ P(|\nu_\mu\rangle \to |\nu_\beta\rangle)
    \label{eq : number of events FD}
\end{equation}
where $\phi_{\nu_\mu}^{FD}$ is the neutrinos flux at the FD, $\sigma_{CC}^{FD}$ is the cross section for charged current interaction at the FD and $P(|\nu_\mu\rangle \to |\nu_\beta\rangle )$ is the corresponding oscillation probability. 
We can conclude that the total number of events $M$ used to compute the Cramér-Rao bound \eqref{eq:qcrb} and corresponding to the {\em non oscillated} neutrinos, is efficiently estimated from the number of events $M_{\nu_\alpha}^{\small ND}$ observed at the ND via the formula 
\begin{align}
 M =  M_{\nu_\mu}^{ND} \biggl(\frac{L_{ND}}{L_{FD}}\biggl)^2 \ \frac{n_{targ}^{ND}}{n_{targ}^{\small FD}} \frac{\sigma_{CC}^{ND}}{\sigma_{CC}^{FD}} 
\end{align}
where we have taken into account the different number of targets and cross sections at the two detectors and that the flux at the FD is linked to the one at the ND as $\phi_{\nu_\mu}^{FD}/ \phi_{\nu_\mu}^{\small ND}\approx (L_{ND}/L_{FD})^2$ given the decrease of the flux with $L^{-2}$. The ratio between the cross sections is equal to one if the ND and the FD are of the same type. \\
For reactor experiments the situation depends on whether a ND is present in the experiment. In Daya Bay a ND is indeed present, and one can follow the arguments above, by simply considering an initial beam of electron antineutrinos in Eq.~\eqref{eq : eventi ND}. As no ND is present in KamLAND, in order to infer the number of total events $M$ one has to resort to theoretical simulations \cite{PhysRevLett.94.081801, PhysRevLett.100.221803,PhysRevLett.90.021802}.\\

Finally, for acceleator experiments we have to address how to consider in a comprehensive way the information obtained by both neutrinos and antineutrinos beams. As the corresponding events can be safely considered as statistically independent, the ultimate bound on the estimator variance ${\rm Var}_{\tilde{\lambda}}(\lambda)$ in Eq.~\eqref{eq:qcrb} taking into account both beams is obtained by replacing the denominator with:
\begin{align}
M H_{\nu,\overline{\nu}}(\lambda) = M (\xi_\nu H_{\nu}(\lambda) + \xi_{\bar\nu} H_{\bar\nu}(\lambda)) \,,
\end{align}
where $\xi_\nu = M_\nu / M$ and $\xi_{\bar\nu} = M_{\bar\nu}/M$ are, respectively, the fractions of the total number of experiments $M=M_\nu + M_{\bar\nu}$ corresponding to neutrino ($M_{\nu}$) or antineutrino ($M_{\bar\nu}$) events that can be estimated at the ND as described above. Our calculations show a-posteriori that, in the parameters regime of interest, $H_{\nu}(\lambda)$ is approximately equal to $H_{\bar\nu}(\lambda)$. Consequently, we can safely quantify the information contained both in neutrino and antineutrino quantum states simply as the average of the individual QFIs, $H_{\nu,\overline{\nu}}(\lambda) = (H_{\nu}(\lambda)+H_{\bar\nu}(\lambda))/2$, and take this as our main figure of merit.
Notice that the same argument will be also applied to the classical FIs corresponding to the measurement strategies we will consider: in fact one can verify the when focusing on the CP-phase $\delta_{CP}$, the neutrino and antineutrino FIs values $F_\nu(\delta_{CP})$ and  $F_{\bar\nu}(\delta_{CP})$ may have a non negligible difference in some regimes, in particular for values of $\delta_{CP}$ corresponding to maximum CP-violation. However, it is still true that the two quantities are approximately equal for values of $L/E$ corresponding to probability oscillation maxima, that is in correspondence of the current and future neutrino experiments. Because of this and since no such difference is observed for the other parameters, we can thus define also the effective FI that takes into account both neutrinos and antineutrinos events for all the parameters as the arithmetic average $F_{\nu,\overline{\nu}}(\lambda) =( F_{\nu}(\lambda)+F_{\bar\nu}(\lambda))/2$. 

These results, which show that combining neutrino and antineutrino experiments does not enhance the amount of information on the CP phase $\delta_{CP}$ compared to focusing on either neutrino or antineutrino experiments alone, may seem confusing. In fact, comparing the oscillation probabilities from both experiments is crucial for the task of discriminating between CP-conserving and CP-violating scenarios, which is, however, fundamentally different from estimating the value of the CP phase $\delta_{CP}$.

As for reactor experiments only electron antineutrinos are involved, we will simply denote respectively with $H_{\bar\nu_e}(\lambda)$ and $F_{\bar\nu_e}(\lambda)$ the corresponding QFI and FI antering in the corresponding Cram\'er-Rao bounds.
\section{Quantum Fisher information for the neutrino oscillation parameters}\label{Quantum Fisher Information}
In this Section, we present the results for the QFI $H_{\nu,\overline{\nu}}(\lambda)$ for accelerator experiments and $H_{\overline{\nu}_e}(\lambda)$ for reactor experiments, considering all the parameters of interest. We will start by discussing the results for the CP-phase $\delta_{CP}$, moving then to the mixing angles $\theta_{jk}$. 
\subsection{QFI for the CP phase $\delta_{CP}$}
\begin{figure}[h]
    \centering
    \includegraphics[width=0.6\textwidth]{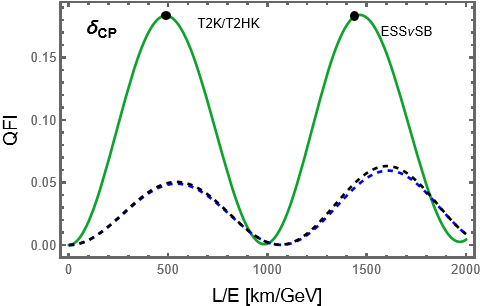}
    \caption{QFI $H_{\nu,\overline{\nu}}(\delta_{CP})$ obtained with a $\nu_\mu / \overline{\nu}_\mu$ beam (green solid line) as a function of the baseline/energy ratio $L/E$. The dashed black and blue curves show the appearance probabilities $P(\nu_{\mu }\to \nu_e)$ and $P(\bar{\nu}_{\mu} \to \bar{\nu}_e)$, respectively. For T2K/T2HK we considered $L=295$ km and a peak energy of 0.6 GeV, while for ESS$\nu$SB $L=360$ km and a peak energy of 0.25 GeV.}
    \label{fig: qfi delta mu}
\end{figure}

At present, $\delta_{CP}$ is probed exclusively in accelerator-based experiments, and thus we here evaluate the QFI $H_{\nu,\overline{\nu}}(\delta_{CP})$ for muon neutrino/antineutrino beams.
In Fig.~\ref{fig: qfi delta mu}, we show the behavior of the QFI $H_{\nu,\overline{\nu}}(\delta_{CP})$ for a $\nu_{\mu}/\bar \nu_{\mu}$ beam. In the range of L/E accessible to current experiments, $H_{\nu,\overline{\nu}}(\delta_{CP})$ exhibits no appreciable dependence on $\delta_{CP}$, implying that the information content of the quantum state is effectively independent of the precise value of $\delta_{CP}$ in this regime. The peak amplitudes remain constant across the explored range of $L/E$, indicating that the information encoded in the state is preserved over successive oscillation maxima. The maxima of the QFI coincide with those of the appearance probability $P(\nu_\mu(\overline{\nu}_\mu)\to\nu_e(\overline{\nu}_e))$, consistent with the fact that maximal appearance enhances sensitivity to $\delta_{CP}$. 

Using Eq.~\eqref{eq:qcrb}, we compute the intrinsic quantum bound at the peak energies of the T2K and ESS$\nu$SB experiments.
Assuming a conservative estimate of a total of $M=\mathcal{O}(10^3)$ events, the resulting bound on the precision of $\delta_{CP}$ is approximately $3^\circ$. 
This indicates that, even with a limited event statistics comparable to current experimental conditions, the achievable precision is not constrained by any fundamental quantum limit. Therefore, the presently limited precision in the measurement of $\delta_{CP}$ cannot be attributed to an intrinsic quantum restriction. \par
\begin{figure}[h]
    \centering
    \includegraphics[width=0.6\textwidth]{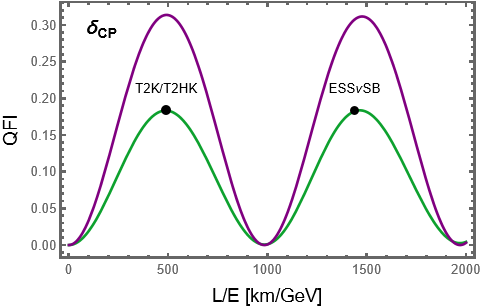}
    \caption{QFI $H_{\nu,\overline{\nu}}(\delta_{CP})$ obtained with a $\nu_\mu / \overline{\nu}_\mu$ beam (green solid line) and QFI $H_{\overline{\nu}_e}(\delta_{CP})$ obtained with a $\nu_e$ beam (purple solid line) as a function of the ratio $L/E$ between baseline and beam energy. For T2K/T2HK we considered $L=295$ km and a peak energy of 0.6 GeV, while for ESS$\nu$SB $L=360$ km and a peak energy of 0.25 GeV.}
    \label{fig:delta_e}
\end{figure}

Although no current or planned experiment aims to measure $\delta_{CP}$ using electron neutrino beams, we have nevertheless computed the QFI for such a beam, as shown in Fig.~\ref{fig:delta_e}.
As illustrated in Fig.~\ref{fig:delta_e}, the information associated with $\delta_{CP}$ in the $\bar\nu_e$ state is slightly larger than that in the $\nu_\mu$ or $\bar{\nu}_\mu$ states. This observation, however, remains speculative, since present and near-future facilities cannot probe the appearance channel with electron neutrino beams in the relevant energy range. Indeed, existing electron neutrino experiments are reactor-based and operate at energies too low to access the appearance channel.
In contrast, neutrino factories—or, more generally, facilities employing muon-decay neutrino beams such as the proposed MOMENT \cite{MOMENT} experiment—can produce high-intensity, high-energy electron neutrino beams that are unattainable with conventional sources~\cite{neutrinoFpilar}. The MOMENT facility~\cite{MOMENT} proposes a novel approach based on the decay of positive and negative muons, enabling the simultaneous study of neutrino and antineutrino oscillations across eight distinct channels.
\subsection{QFI for mixing angles $\theta_{12}$, $\theta_{13}$ and $\theta_{23}$}

\begin{figure}[h]
    \centering
    \includegraphics[width=0.6\textwidth]{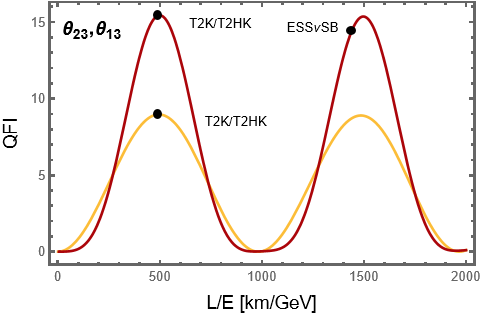}
    \caption{QFI for $\theta_{23}$ (bordeaux) and $\theta_{13}$ (orange) for a muon (anti)neutrino beam. For T2K/T2HK we considered $L=295$ km and a peak energy of 0.6 GeV, while for ESS$\nu$SB $L=360$ km and a peak energy of 0.25 GeV.}
    \label{fig:23_mu}
\end{figure}

As discussed above, accelerator-based experiments aim to estimate the mixing angles $\theta_{13}$ and $\theta_{23}$. We thus start by discussing the QFIs $H_{\nu,\overline{\nu}}(\theta_{13})$ and $H_{\nu,\overline{\nu}}(\theta_{23})$ for muon (anti)neutrino beams and display their behaviour in Fig.~\ref{fig:23_mu} as a function of $L/E$. As for $\delta_{CP}$, the QFI is effectively independent of $\delta_{CP}$ within the experimentally relevant range, and for clarity only the case $\delta_{CP}^{(\mathrm{NO})}$ is shown.  Consistent with expectations, $H_{\nu,\overline{\nu}}(\theta_{13})$ and $H_{\nu,\overline{\nu}}(\theta_{23})$ reach minima when the survival probability is maximal, $P(\nu_\mu \to \nu_\mu) \approx 1$, corresponding to an absence of oscillations and, consequently, minimal information about these parameters. The maxima of these QFIs are approximately $H_{\nu,\overline{\nu}}(\theta_{13}) \simeq 8$ and $H_{\nu,\overline{\nu}}(\theta_{23}) \simeq 15$, which are, roughly, an order and two orders of magnitude larger, than the corresponding value for $\delta_{CP}$, respectively. This implies that, for the same number of events, the attainable uncertainty on $\theta_{13}$ or $\theta_{23}$ are respectively about a factor 3 and an order of magnitude smaller than that on $\delta_{CP}$ .

\par
\begin{figure}[h]
    \centering
    \begin{subfigure}{0.45\textwidth}
        \includegraphics[width=\textwidth]{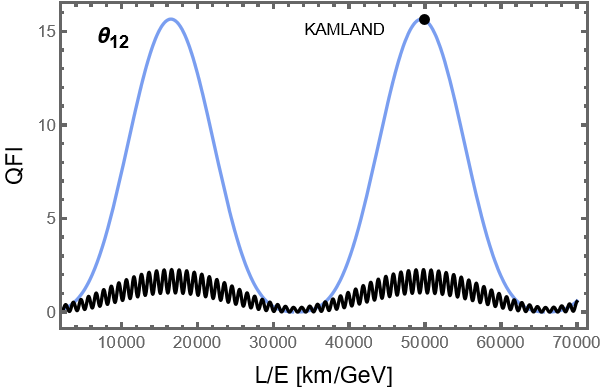}\caption{}
        \label{fig:12_e}
    \end{subfigure}
    \begin{subfigure}{0.45\textwidth}
        \includegraphics[width=\textwidth]{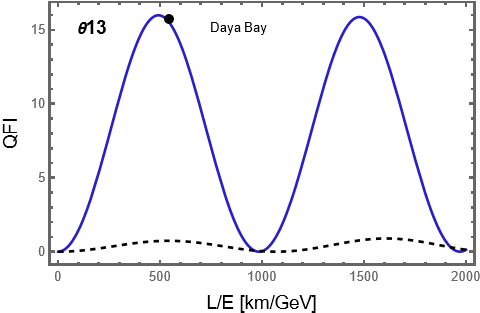}\caption{}
        \label{fig:13_e}
    \end{subfigure}

    \caption{QFI $H_{\bar{\nu}_e}(\theta_{12})$ (a)  and $H_{\bar{\nu}_e}(\theta_{13})$ (b) for an electron antineutrino beam. In black the oscillation probability $P(|\overline{\nu}_e\rangle  \to |\overline{\nu}_\mu\rangle)$ (multiplied by 5 in (a) and by 15 in (b)). For KamLand $L=180 $km and the peak energy lies between 3-4 MeV and we choose $E=3.6$ MeV. For Daya Bay $L=1.64$ km and we considered a peak energy of 3 MeV.}
\end{figure}

We have then considered the estimation of $\theta_{12}$ and $\theta_{13}$ in reactor-based experiments, and thus evaluated the QFI $H_{\bar{\nu}_e}(\theta_{12})$ and $H_{\bar{\nu}_e}(\theta_{13})$, for an electron antineutrino beam. As shown in Fig.~\ref{fig:12_e}, $H_{\bar{\nu}_e}(\theta_{12})$ exhibits sensitivity to the slow oscillations governed by $\Delta m_{21}^2$. As in the accelerator-based cases, the QFI reaches its maxima when the antineutrino has a non-zero probability of oscillating into another flavor and its minima when the survival probability approaches unity. The peak value, $H_{\bar{\nu}_e}(\theta_{12}) \approx 15$, is also substantially larger than that obtained for $\delta_{CP}$.

Hence, a key observation that emerges from these results is that $H_{\nu,\bar{\nu}}(\delta_{CP})$ is roughly an order of magnitude smaller than the QFI for all mixing angles. In other words, the quantum states $|\nu_\mu(t)\rangle$, $|\bar{\nu}_\mu(t)\rangle$ and $|\bar{\nu}_e(t)\rangle$ carry significantly less information about $\delta_{CP}$ than about $\theta_{23}$, $\theta_{13}$, or $\theta_{12}$. Among these, $H_{\nu,\bar{\nu}}(\theta_{23})$, $H_{\bar{\nu}_e}(\theta_{12})$ and $H_{\bar{\nu}_e}(\theta_{13})$ attain the largest values, indicating that the quantum state encodes the most information about $\theta_{23}$, $\theta_{12}$ and $\theta_{13}$ thus sets a lower fundamental bound on the variance compared to  $\delta_{CP}$.
This result quantitatively demonstrates that, at a fundamental level, estimating the CP-violating phase is intrinsically more challenging than determining the other PMNS mixing parameters. As we later discuss in Sec.~\ref{Fisher Information}, this disparity becomes even more pronounced when flavor measurement strategies are taken into account.

\section{Fisher Information for neutrino oscillation parameters with flavor measurements}\label{Fisher Information}
While the QFI provides the ultimate limits imposed by quantum mechanics, actual experiments measure the event rates for different neutrino flavors. Consequently, it is essential to compute the FI associated with such measurement strategies, and in order to do that one first needs to identify the corresponding PVM mathematically describing them. 

In accelerator experiments, detectors are able to detect both electron and muon (anti)neutrinos, corresponding to the so-called appearance and disappearance channels. We start by considering the ideal case in which information is extracted from both channels, such that the corresponding PVM is described by the projectors: 
\begin{equation}
    \Pi_e = |\nu_e\rangle\langle \nu_e|, \quad
    \Pi_\mu = |\nu_\mu\rangle\langle \nu_\mu|, \quad
    \Pi_{\rm off} = \mathbb{I} - \Pi_e - \Pi_\mu = |\nu_\tau\rangle\langle \nu_\tau|.
    \label{eq:flavor_pvm}
\end{equation}
Here, $\Pi_e$ and $\Pi_\mu$ project onto the electron and muon neutrino eigenstates, while $\Pi_{\rm off}$ accounts for all outcomes other than $\nu_e$ or $\nu_\mu$, which correspond exclusively to $\nu_\tau$. Even though $\nu_\tau$ cannot be directly detected, the Hilbert space of possible outcomes remains three-dimensional, and since,  as we mentioned before, we can efficiently estimate the total number of neutrinos $M$ generated and involved in the experiment, the number of tau neutrinos can be inferred as $M_\tau = M - (M_\nu+M_e)$, given the numbers of the detected muon and electron neutrinos $M_\mu$ and $M_e$. 

We will also provide results by considering the scenario where either the appearance or the disappearance channel is considered separately. The PVM for the appearance channel reads
\begin{equation}
   \Pi_e=|\nu_e\rangle\langle\nu_e|,\quad
   \Pi_{\rm off}=\mathbb{I}-\Pi_e,
   \label{eq:appeareancePVM}
\end{equation}
while for the disappearance channel one has
\begin{equation}
   \Pi_\mu=|\nu_\mu\rangle\langle\nu_\mu|,\quad
   \Pi_{\rm off}=\mathbb{I}-\Pi_\mu,
   \label{eq:disappeareancePVM}
\end{equation}
In these cases, the events obtained at the FD will give information only on the neutrinos detected in the corresponding channel, $M_e$ and $M_\mu$ respectively. However, also here, the neutrinos corresponding to the {\em off} results can be obtained respectively as $M_{\sf off} = M - M_e$ and $M_{\rm off} = M - M_\mu$ for the two channels, once the information on the total number of neutrinos $M$ is estimated as described in Sec.~\ref{s:numberevents}.

The same construction applies for a $\overline{\nu}_\mu$ beam, obtaining the corresponding PVM by using the antineutrino flavor projectors.\\

\par
As in reactor experiments only the disappearance channel $\overline{\nu}_e \to \overline{\nu}_e$ is probed, we have just considered the corresponding PVM, which reads
\begin{equation}
    \Pi_{e^+} = |\overline{\nu}_e\rangle\langle\overline{\nu}_e|,
\end{equation}
\begin{equation}
    \Pi_{{\rm off}} = \mathbb{I} - \Pi_{e^+}.
    \label{eq:povm_reactor}
\end{equation}
Given the PVMs just outlined, the corresponding FIs can be readily evaluated via Eq.~\eqref{eq:FI}.

As with the QFI, when evealuting these formulas, we assume the estimation of a single parameter $\lambda$, with all other parameters held fixed. For the fixed parameters and for generating the FI plots, we adopt the NuFit-6.0 best-fit values~\cite{Esteban_2024}, with the exception of $\delta_{CP}$ for accelerator experiments. In these cases, we also consider $\delta_{CP}=0$ and $\pi/2$, reflecting the large uncertainty in its current determination. For reactor experiments, which are insensitive to $\delta_{CP}$, the FI is independent of its value; for convenience, we fix $\delta_{CP}$ to the NuFit-6.0 best-fit value, denoted $\delta_{CP} = \delta_{CP}^{(\mathrm{NO})}$.\par

\subsection{FI for the CP phase $\delta_{CP}$}
\label{FI delta}
 \begin{figure}[h]
    \centering
    \begin{subfigure}{0.45\textwidth}
        \includegraphics[width=\textwidth]{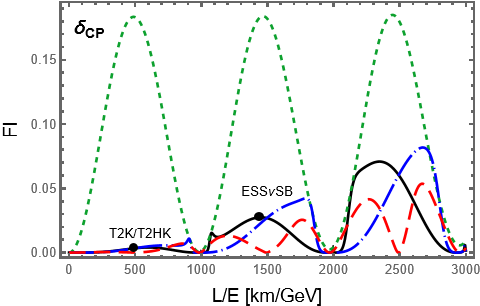}\caption{}
        \label{fig:comparison_mu}
    \end{subfigure}
    \begin{subfigure}{0.45\textwidth}
        \includegraphics[width=\textwidth]{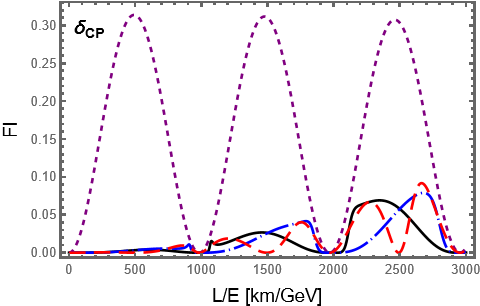}\caption{}
        \label{fig:comparison_e_delta}
    \end{subfigure}

    \caption{Comparison of the FI $F_{\nu,\overline{\nu}}(\delta_{CP})$ evaluated at $\delta_{CP}=\delta_{CP}^{(\mathrm{NO})}$ (solid black line), $\delta_{CP}=0$ (dash dotted blue line) and $\delta_{CP}=\pi/2$ (dashed red line) for a muon (anti)neutrino beam (a) and for an electron antineutrino beam (b). The corresponding QFI $H_{\nu,\overline{\nu}}(\delta_{CP})$ is reported as a dashed green line and $H_{\overline{\nu}_e}(\delta_{CP})$ as a dashed purple line.  For T2K/T2HK we considered $L=295$ km and a peak energy of 0.6 GeV, while for ESS$\nu$SB $L=360$ km and a peak energy of 0.25 GeV.}
\end{figure}
We start by considering the FI $F_{\nu,\overline{\nu}}(\delta_{CP})$ corresponding to the ideal flavor measurement \eqref{eq:flavor_pvm} on muon neutrino and antineutrino beams in accelerator experiments.
Figure~\ref{fig:comparison_mu} shows $F_{\nu,\overline{\nu}}(\delta_{CP})$ compared with the QFI $H_{\nu,\overline{\nu}}(\delta_{CP})$ as a function of $L/E$.
We observe that, unlike the QFI, the FI $F_{\nu,\overline{\nu}}(\delta_{CP})$ depends both on the value of $\delta_{CP}$ and on the oscillation peak considered. As noted in~\cite{prl}, $F_{\nu,\overline{\nu}}(\delta_{CP})$ never reaches the peaks of $H_{\nu,\overline{\nu}}(\delta_{CP})$, remaining significantly lower throughout. Since the FI depends on the measurement strategy, this indicates that the currently implemented strategy, being the only feasible one, is not optimal for measuring $\delta_{CP}$.
A strategy based on operating at the first oscillation maximum appears to be strongly suboptimal, as it yields FI values that are several orders of magnitude smaller than the corresponding QFI. In contrast, the FI is typically enhanced in the vicinity of the second oscillation maximum. 
Here, we consider also electron antineutrino beams and we notice a similar behaviour, that is FI does not saturate the QFI peaks and exhibits  an enhancement at the second oscillation peak as shown in Fig. \ref{fig:comparison_e_delta} .
 \begin{figure}[h]
    \centering

    \begin{subfigure}{0.45\textwidth}
        \centering
        \includegraphics[width=\textwidth]{immagini/F2F1.png}
       \caption{}
        
    \end{subfigure}
    \hfill
    \begin{subfigure}{0.45\textwidth}
        \centering
        \includegraphics[width=\textwidth]{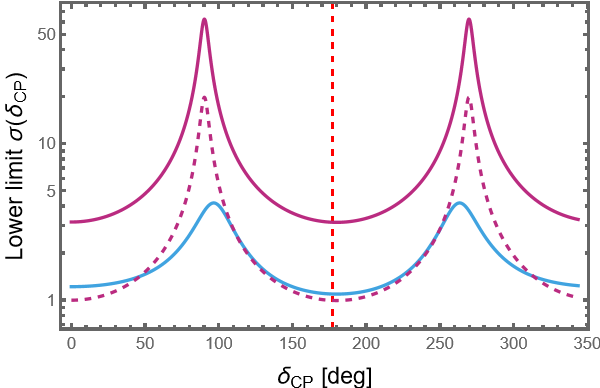}
        \caption{}

    \end{subfigure}

     \caption{(a) Logarithmic plot of the square root of the ratio between the FI at the second (ESS$\nu$SB) and first peak (T2K/T2HK), plotted as a function of $\delta_{CP}$. 
    (b) Logarithmic plot of the lower bound $\sigma(\delta_{CP}) = (M F_{\nu,\overline{\nu}}(\delta_{CP}))^{-1/2}$ on the standard deviation of $\delta_{CP}$ for ESS$\nu$SB, considering $M=10^5$ events \cite{Alekou2022} (blue solid line), for an experiment positioned at the first peak with the same statistics (pink solid line), and pink dashed, considering the same beam power of ESS$\nu$SB but for an experiment positioned at the first peak resulting in approximately $M=10^6$ (pink dashed line), plotted as a function of $\delta_{CP}$. The dashed red line in both plots corresponds to $\delta_{CP}=\delta_{CP}^{(\mathrm{NO})}$.
    }
    \label{peak1peak2}
\end{figure}
Let us now examine in greater detail the behavior of the FI at the first and second oscillation peaks.
In Fig.~\ref{peak1peak2} (panel a) we show the ratio of the FI at the first and second oscillation peaks. The ratio is highly sensitive to the value of $\delta_{CP}$, particularly near maximal CP violation.  

In Fig.~\ref{peak1peak2} (panel b) we compare the maximal achievable precision, $\sigma$, derived from the Cramér–Rao bound in Eq.~\eqref{cramerrao}.
We consider ESS$\nu$SB with $M=10^5$ events \cite{Alekou2022}, and a hypothetical experiment operating at the first oscillation maximum with the same statistics as ESS$\nu$SB first and then with the same beam power as ESS$\nu$SB. The advantage over the experiment at the first peak is evident across the entire range of $\delta_{CP}$.

We find that ESS$\nu$SB systematically achieves a slightly better precision than configurations at the first maximum, even when the latter are assumed to accumulate an order of magnitude larger statistics. The improvement is particularly pronounced in the vicinity of maximal CP violation. 
\par
Our quantum metrology approach validates the conclusions drawn from previous oscillation probability analyses~\cite{pilar} regarding the optimal oscillation peak. 
From an information-theoretic perspective, flavor measurements are intrinsically sub-optimal, particularly at the first oscillation maximum. We demonstrate that this limitation is not merely a feature of the probability-level description, but rather reflects fundamental bounds identified within QET. In this sense, the previously observed advantage of the second oscillation maximum~\cite{pilar} is elevated from a phenomenological observation to a consequence of quantum limits.
To clarify the scope of this analysis, the quantum and classical Fisher information study should be understood as complementary diagnostic tools for mapping out information content and measurement optimality. They are not intended to substitute for standard, full-scale experimental sensitivity fits, but rather to provide a clean, information-theoretic baseline for the system's behavior.

\begin{figure}[h]
    \centering
    \includegraphics[width=0.6\textwidth]{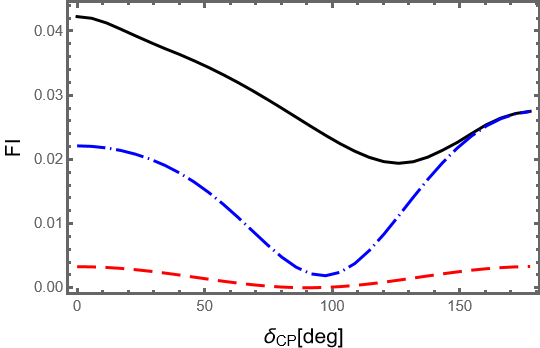}
    \caption{Classical effective Fisher information for flavor detection, $F_{\nu,\overline{\nu}}(\delta_{CP})$, as a function of $\delta_{CP}$ for three representative values of the effective baseline $L/E$. The black solid curve corresponds to the FI optimized over the first two oscillation periods of the neutrino probability; the red long-dashed curve is evaluated at the first oscillation peak; and the blue dash-dotted curve at the second oscillation peak.}
    \label{fig:FIvsdelta}
\end{figure}

Our analysis also reveals nontrivial features beyond earlier studies: the possible enhancement at the second peak depends in fact sensitively on the actual value of $\delta_{CP}$. To further investigate this behaviour in Fig.~\ref{fig:FIvsdelta} we have plotted $F_{\nu,\overline{\nu}}(\delta_{CP})$ as a function of $\delta_{CP}$ for three different values $L/E$: at the first oscillation peak, at the second peak, and optimized over the first two oscillation periods. We observe that the second peak  typically yields larger FI than the first. However, this enhancement is significant only in specific regions of parameter space: it is most pronounced for large $\delta_{CP}$, particularly near $\delta_{CP}^{\mathrm{(NO)}}$, and strongly suppressed around $\delta_{CP}=\pi/2$, where the FI is maximized away from the oscillation peaks. 
This illustrates the potential for “information blind spots’’ in fixed-beam setups. Tunable configurations—such as off-axis sampling (e.g., DUNE-PRISM) or adjustable horn currents—allow dynamic optimization of the flux peak.Our results indicate that optimizing the beam energy within the 200–250 MeV window could enhance the overall information content. A more thorough exploration of this is deferred to future work.

\begin{figure}[h]
    \centering
    \includegraphics[width=0.6\textwidth]{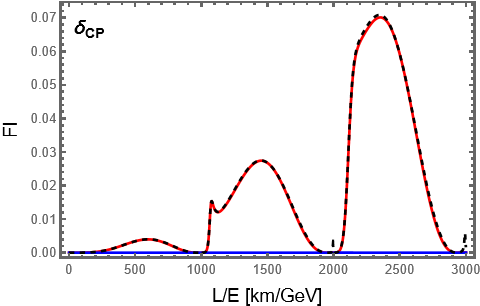}
    \caption{$F_{\nu,\overline{\nu}}(\delta_{CP})$ for $\delta_{CP}=\delta_{CP}^{(\mathrm{NO})}$ and as a function of $L/E$ obtained considering the idealized three flavor measurements (black dashed), the appearance channel (red) and the disappearance channel (blue). }
    \label{fig:comparison_app_dis}
\end{figure}
So far, we have considered an idealized flavor measurement strategy, which allows to discriminate between all the three different possible flavors of the neutrino quantum state. In practice, the PMNS parameters are inferred from either the appearance ($\nu_\mu(\overline{\nu}_\mu) \to \nu_e(\overline{\nu}_e)$) or disappearance ($\nu_\mu(\overline{\nu}_\mu) \to \nu_\mu(\overline{\nu}_\mu)$) channels, which are mathematically described by the PVMs in Eqs.~\eqref{eq:appeareancePVM} and~\eqref{eq:disappeareancePVM}. It is therefore instructive to compare the FI in these realistic scenarios with the idealized case.   
Figure~\ref{fig:comparison_app_dis} shows $F_{\nu,\overline{\nu}}(\delta_{CP})$ for appearance and disappearance channels as a function of $L/E$. As expected, the FI in the appearance channel is significantly larger, while it is nearly zero in the disappearance channel. Remarkably the appearance-channel FI closely matches the idealized flavor measurements FI, highlighting the fact that, if only flavor information is available, the appearance channel is indeed optimal and one does not need to resolve all three neutrino flavors in order to achieve the best sensitivity.  

\subsubsection{Non monochromatic spectrum}
\begin{figure}[h]
    \centering
    \includegraphics[width=0.6\textwidth]{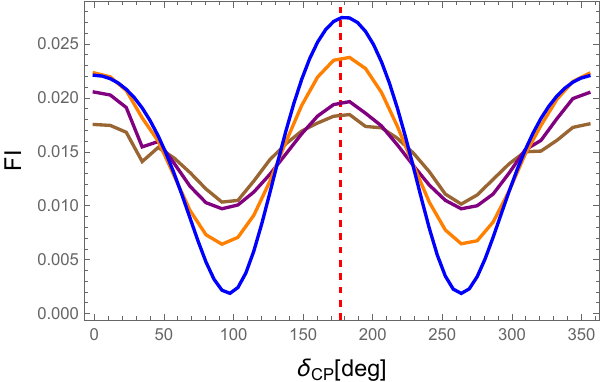}
    \caption{$F_{\nu,\overline{\nu}}(\delta_{CP})$ as a function of $\delta_{CP}$, obtained by considering the peak energy of the ESS$\nu$SB spectrum, $\bar{E}=0.25$ GeV (blue solid line), and an energy spectrum approximated by a Gaussian distribution $\mathcal{G}_{\bar{E},\sigma}(E)$ with mean equal to the peak energy $\bar{E}$ and three different values of the standard deviation: $\sigma = 0.08$ GeV (brown), $\sigma = 0.04$ GeV (purple), and $\sigma = 0.02$ GeV (orange).}
    \label{fig:spettro}
\end{figure}

In our analysis, we derived the QFI and FI by considering only the peak energy of the spectrum for each experiment. In Fig.~\ref{fig:spettro}, we illustrate how the FI changes when the energy spectrum is approximated by a Gaussian distribution. 
In more detail, we consider Gaussian neutrino energy spectra $\mathcal{G}_{\bar{E},\sigma}(E)$ centered at the ESS$\nu$SB peak energy $\bar{E}$, with different values of the standard deviation $\sigma$ (reported in the caption of Fig.~\ref{fig:spettro}). We assume a total of $10^5$ events distributed over 30 energy bins, each one centered at an energy $E_j$ and width $\Delta E$. For each energy $E_j$, we have evaluated the corresponding FI $F_{E_j}(\delta_{CP})$. As the probability $\mathcal{G}_{\bar{E},\sigma}(E_j)\Delta E$ determines the fraction of events falling within the corresponding energy interval, and since events occurring at different energies are effectively independent, we can exploit the additivity of the FI and evaluate the overall effective FI as $F_{\bar{E},\sigma}(\delta_{CP}) =
\sum_j F_{E_j}(\delta_{CP}) \, \mathcal{G}_{\bar{E},\sigma}(E_j)\Delta E$.

\par
When the FI is already close to its maximum at the peak energy, integrating over the full spectrum tends to reduce its value. Conversely, if the peak energy does not coincide with the region of maximal sensitivity, a full spectral analysis can increase the FI and modify the quantitative comparison.
Nevertheless, these effects do not qualitatively affect our main conclusion, namely the existence of a significant gap between the information encoded in the neutrino quantum state and the information extracted through flavor measurements.
\subsection{FI for the mixing angles $\theta_{12}, \theta_{13}$ and $\theta_{23}$}
\begin{figure}[h]
    \centering

    \begin{subfigure}{0.45\textwidth}
        \centering
        \includegraphics[width=\textwidth]{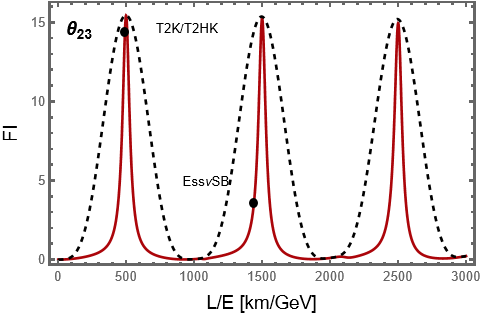}
        \caption{}
        \label{FItheta23}
    \end{subfigure}
    \hfill
    \begin{subfigure}{0.45\textwidth}
        \centering
        \includegraphics[width=\textwidth]{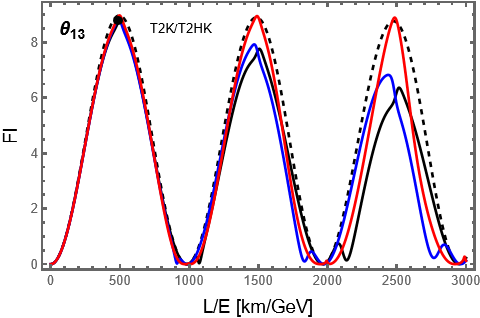}
        \caption{}
        \label{FItheta13}
    \end{subfigure}

    \caption{(a) FI for idealized flavor measurement and muon (anti)neutrino beam $F_{\nu,\overline{\nu}}(\theta_{23})$ (red line) compared to the corresponding QFI $H_{\nu,\overline{\nu}}(\theta_{23})$ (black dashed) as a function of $L/E$ and evaluated at $\delta_{CP}=\delta_{CP}^{(\mathrm{NO})}$. (b)
    FI for idealized flavor measurement and muon (anti)neutrino beam $F_{\nu,\overline{\nu}}(\theta_{13})$ as a function of $L/E$ evaluated at $\delta_{CP}=\delta_{CP}^{(\mathrm{NO})}$ (black solid line), at $\delta_{CP}=0$ (blue solid line) and $\delta_{CP}=\pi/2$ (red solid line). The black dashed line shows the QFI $H_{\nu,\overline{\nu}}(\theta_{13})$ evaluated at $\delta_{CP}=\delta_{CP}^{(\mathrm{NO})}$.  For T2K/T2HK we considered $L=295$ km and a peak energy of 0.6 GeV, while for ESS$\nu$SB $L=360$ km and a peak energy of 0.25 GeV.}
    
\end{figure}

\begin{figure}[h]
    \centering

    \begin{subfigure}{0.45\textwidth}
        \centering
        \includegraphics[width=\textwidth]{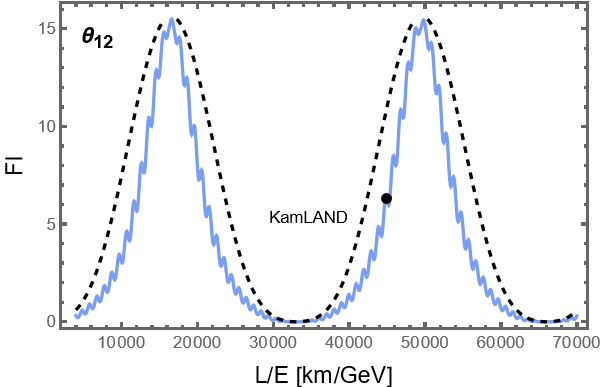}
        \caption{}
        \label{FItheta12}
    \end{subfigure}
    \hfill
    \begin{subfigure}{0.45\textwidth}
        \centering
        \includegraphics[width=\textwidth]{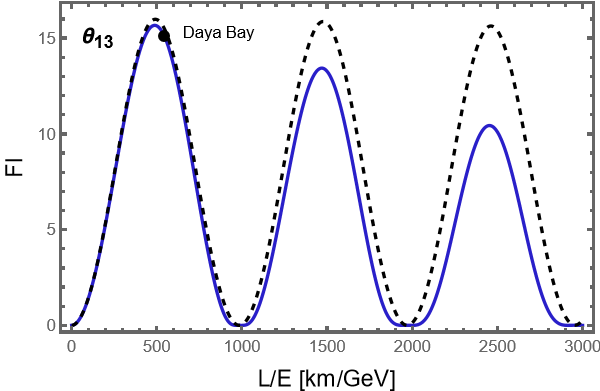}
        \caption{}
        \label{FItheta13e}
    \end{subfigure}

    \caption{Plots obtained for an electron antineutrino beam in reactor experiments. (a) In light blue $F_{\overline{\nu}_e}(\theta_{12})$, in black dashed $H_{\overline{\nu}_e}(\theta_{12})$ evaluated at $\delta_{CP}=\delta_{CP}^{(\mathrm{NO})}$. (b) In blue $F_{\overline{\nu}_e}(\theta_{13})$, in black dashed $H_{\overline{\nu}_e}(\theta_{13})$ evaluated at $\delta_{CP}=\delta_{CP}^{(\mathrm{NO})}$. }
    
\end{figure}
We start by addressing the estimation precision achievable for the mixing angles $\theta_{23}$ and $\theta_{13}$ via muon (anti)neutrino beams in accelerator experiments via idealized flavor measurement.
Figure~\ref{FItheta23} compares the FI, $F_{\nu,\overline{\nu}}(\theta_{23})$, with the corresponding QFI, $H_{\nu,\overline{\nu}}(\theta_{23})$, for a muon (anti)neutrino beam, while Figure~\ref{FItheta13} presents the analogous comparison for $\theta_{13}$.
Contrarily to what we have observed for the CP-phase $\delta_{CP}$, we here observe that the flavor measurement is already optimal, coinciding with the QFI, at the first peak for both angles $\theta_{23}$ and $\theta_{13}$.
While for $\theta_{23}$ this beaviour is confirmed also at the following oscillation peaks, for $\theta_{13}$, a residual dependence on $\delta_{CP}$ appears beyond the first oscillation peak; however, this effect is not phenomenologically relevant, as $\theta_{13}$ is primarily determined at the first oscillation maximum, where the FI saturates the QFI bound. Consequently, T2K and T2HK provide optimal sensitivity to $\theta_{13}$ and $\theta_{23}$, while ESS$\nu$SB is suboptimal for $\theta_{23}$ due to its lower peak energy.

We have then addressed the estimation precision for angles $\theta_{12}$ and $\theta_{13}$ obtained via electron antineutrino beams in reactor experiments. The values of the FIs are presented in Figs.~\ref{FItheta12} and \ref{FItheta13e}, along with the corresponding QFI. We find that $F_{\bar{\nu}_e}(\theta_{12})$ is sensitive to both fast and slow oscillation components, while $F_{\bar{\nu}_e}(\theta_{13})$ responds primarily to the fast mode. Moreover, $F_{\bar{\nu}_e}(\theta_{13})$ decreases as the antineutrino propagates, its value at the second oscillation maximum is smaller than at the first. At the first maximum, corresponding to the energy range probed by Daya Bay \cite{An_2014}, the FI equals the QFI, confirming that the measurement strategy is optimal and that Daya Bay is ideally suited for determining $\theta_{13}$. Conversely, $F(\theta_{12})$ remains stable and reaches the QFI maximum, indicating that KamLAND provides an equally optimal configuration for measuring $\theta_{12}$ at its characteristic energy.

Concluding, for the mixing angles, we find that the flavor measurement strategy is optimal, saturating the QFI at the first peak of the neutrino oscillations. In contrast, as shown in the previous section, for $\delta_{CP}$ there remains a significant gap between the FI and the QFI.

\section{Discussion and Outlook} \label{conclusions}
In this work, we have explored the ultimate precision limits of neutrino oscillation measurements from a quantum metrology perspective. We addressed two fundamental questions: first, what is the intrinsic quantum bound on experimental precision as determined by the QFI, and second, whether the only experimentally feasible strategy—flavor measurements—can be interpreted as a practically accessible quantum-limited bound on parameter estimation.
Our results indicate that flavor measurement strategies for determining the mixing angles are close to optimal, capturing nearly all the accessible information in the neutrino quantum state under idealized conditions. The situation for $\delta_{\rm CP}$ is different: while we find that the quantum state contains significantly less information on this parameter than on the mixing angles, the limited precision on the CP-violating phase arises primarily not from intrinsic quantum limits, but from the low Fisher information at the first $\nu_\mu \to \nu_e$ oscillation peak. Sensitivity can be substantially enhanced by targeting the second oscillation maximum, as proposed for the ESS$\nu$SB experiment \cite{pilar,Alekou_2023}, however, for certain values of $\delta_{CP}$, an alternative strategy may be envisaged. For example, one could explore tuning the effective baseline $L/E$ to probe specific regions of the $\delta_{CP}$ parameter space, or exploiting flexible-spectrum configurations such as DUNE-PRISM to enhance the achievable sensitivity. We leave this investigation to future work.
We further find that electron antineutrino beams carry more information on $\delta_{\rm CP}$ than muon (anti)neutrino beams, and that focusing on the second oscillation peak may be a promising approach. This highlights future neutrino factories and high-energy muon-decay experiments as particularly suitable platforms for precision studies of CP violation. Overall, our work demonstrates that the principal path to improved $\delta_{\rm CP}$ sensitivity lies in the strategic design of experiments, especially through the selection of higher oscillation maxima.
The next step is to incorporate realistic experimental effects to fully exploit quantum metrology as a tool for experimental design. These include energy-dependent detection efficiencies, matter effects in long-baseline experiments such as DUNE \cite{dune,universe10050221} and NO$\nu$A \cite{PhysRevLett.123.151803}, modeling mass eigenstates as wave packets to include decoherence effects, and extending the analysis to multiparameter estimation for the simultaneous measurement of several oscillation parameters.\\

\noindent

{\em Acknowledgements} The authors acknowledge insightful discussions with P. Coloma and A. Serafini, and in particular thank F. Terranova for his valuable help and stimulating discussions.
\\

\noindent

{\em Note added} - During the final stages of this work, two related preprints \cite{Yadav:2026lsx, Chundawat:2026jjd} appeared. In particular, Ref.~\cite{Yadav:2026lsx} focuses exclusively on the QFI of various oscillation parameters, whereas Ref.~\cite{Chundawat:2026jjd} studies the gap between the classical FI information and the QFI within a simplified framework of two-flavor oscillations relevant for reactor and solar neutrino experiments.
\appendix 

\thispagestyle{empty}
\mbox{} 
\bibliographystyle{JHEP}    
\bibliography{biblio}  
\end{document}